\documentclass[english]{article}%
\usepackage{amssymb}
\usepackage{fontenc}
\usepackage[latin1]{inputenc}
\usepackage{amsmath}
\usepackage{fontenc}
\usepackage{babel}
\usepackage{amsfonts}
\usepackage{graphics}
\usepackage{graphicx}%
 
\ifx\pdfoutput\relax\let\pdfoutput=\undefined\fi
\newcount\msipdfoutput
\ifx\pdfoutput\undefined\else
\ifcase\pdfoutput\else
\ifx\paperwidth\undefined\else
\ifdim\paperheight=0pt\relax\else\pdfpageheight\paperheight\fi
\ifdim\paperwidth=0pt\relax\else\pdfpagewidth\paperwidth\fi
\fi\fi\fi
\begin{document}

\title{{\Large Coherent states of systems with quadratic Hamiltonians}}
\author{V.G. Bagrov\thanks{Department of Physics, Tomsk State University, Russia;
Institute of High Current Electronics, SB RAS, Tomsk, Russia; e-mail:
bagrov@phys.tsu.ru }, D. M. Gitman\thanks{Tomsk State University, Russia; P.N.
Lebedev Physical Institute, Russia; Institute of Physics, University of São
Paulo, Brazil; e-mail: gitman@if.usp.br}, and A. S. Pereira\thanks{Institute
of Physics, University of São Paulo, Brazil; e-mail: apereira@if.usp.br}}
\maketitle

\begin{abstract}
Different families of generalized CS for one-dimensional systems with general
time-dependent quadratic Hamiltonian are constructed. In principle, all known
CS of systems with quadratic Hamiltonian are members of these families. Some
of the constructed generalized CS are close enough to the well-known due to
Schrödinger and Glauber CS of a harmonic oscillator, we call them simply CS.
However, even among these CS there exist different families of complete sets
of CS. These families differ by values of standard deviations at the initial
time instant. According to the values of these initial standard deviations one
can identify some of the families with semiclassical CS. We discuss properties
of the constructed CS, in particular, completeness relations, minimization of
uncertainty relations and so on. As a unknown application of the general
construction, we consider different CS of an oscillator with a time dependent frequency.


\end{abstract}

\section{Introduction}

\subsection{General}

Coherent states (CS) play an important role in modern quantum theory as states
that provide a natural relation between quantum mechanical and classical
descriptions. They have a number of useful properties and as a consequence a
wide range of applications, e.g. in semiclassical description of quantum
systems, in quantization theory, in condensed matter physics, in radiation
theory, in quantum computations, in loop quantum gravity, and so on, see, e.g.
Refs. \cite{CSQT72,MalMa,MalMa68,DodMa87}. Despite the fact that there exist a
great number of publications devoted to constructing CS of different systems,
an universal definition of CS and a constructive scheme of their constructing
for arbitrary physical system is not known. However, it seems that for systems
with quadratic Hamiltonians there exist at present a common point of view on
this problem\footnote{In this article we do not discuss the so-called
generalized CS \cite{gilm72,perel72,perel86,AAG00}.}. Starting the works
\cite{MalMa,DodMa87,DodMa03,16} CS are defined as eigenvectors of some
annihilation operators that are at the same time integrals of motion, see also
\cite{4,265,287,290,295,303}. Of course such defined CS have to satisfy the
corresponding Schrödinger equation. In the frame of such a definition one can
in principle construct CS for a general quadratic system. This construction is
based on solutions of some classical equations, their analysis represent a
nontrivial part of the CS construction.

In this article we, following, the integral of motion method, construct
different families of generalized CS for one-dimensional systems with general
time-dependent quadratic Hamiltonian. Analyzing these families, we see that
some of them are more close to the well-known due to Schrödinger and Glauber
CS (see \cite{Glauber}) of a harmonic oscillator, we call them simply CS.
However, among the latter CS there exist still different families of complete
sets of CS. These families differ by values of standard deviations at the
initial time instant. According to the values of these initial standard
deviations one can identify some of the families with semiclassical CS, as was
demonstrated by us in the free particle case \cite{BagGiP14}. We discuss
properties of the constructed CS, in particular, completeness relations,
minimization of uncertainty relations and so on. As an application of the
general construction, we consider CS of an oscillator with a time dependent frequency.

\subsection{Basic equations\label{S2}}

Consider quantum motion of a one-dimensional system with the generalized
coordinate $x$ on the whole real axis, $x\in\mathbb{R}=\left(  -\infty
,\infty\right)  ,$ supposing that the corresponding quantum Hamiltonian
$\hat{H}_{x}$ is given by a quadratic form of the operator $x$ and the
momentum operator $\hat{p}_{x}=-i\hbar\partial_{x}$,%
\begin{equation}
\hat{H}_{x}=r_{1}\hat{p}_{x}^{2}+r_{2}x^{2}+r_{3}(x\hat{p}_{x}+\hat{p}%
_{x}x)+r_{4}x+r_{5}\hat{p}_{x}+r_{6}, \label{1.1}%
\end{equation}
where $r_{s}=r_{s}\left(  t\right)  ,$ $s=1,...,6$ are some given functions of
the time $t.$ We suppose that these functions are real and both $\hat{H}_{x}$
and $\hat{p}_{x}$ are self-adjoint on their natural domains $D_{H_{x}}$ and
$D_{p_{x}}$ respectively, see e.g. \cite{book2}.

Quantum states of the system under consideration are described by a wave
function $\Psi\left(  x,t\right)  $ which satisfies the Schrödinger equation
\begin{equation}
i\hbar\partial_{t}\Psi\left(  x,t\right)  =\hat{H}_{x}\Psi\left(  x,t\right)
. \label{1.2}%
\end{equation}
In what follows, we restrict ourselves by a physically reasonable case
$r_{1}\left(  t\right)  >0$. In this case, we introduce dimensionless
variables, a coordinate $q$ and a time $\tau$ as follows%
\begin{equation}
q=xl^{-1},\text{ \ }\tau=\int_{0}^{t}\frac{ds}{T\left(  s\right)  }%
=\frac{2\hbar}{l^{2}}\int_{0}^{t}r_{1}\left(  s\right)  ds,\ \ T\left(
t\right)  =\frac{l^{2}}{2\hbar r_{1}\left(  t\right)  }, \label{1.3}%
\end{equation}
where $l$ is an arbitrary constant of the dimension of the length. The new
momentum operator $\hat{p}$ and the new wave function $\psi\left(
q,\tau\right)  $\ read%
\begin{equation}
\hat{p}=\frac{l}{\hbar}\hat{p}_{x}=-i\partial_{q},\ \ \psi\left(
q,\tau\right)  =\sqrt{l}\Psi\left(  lq,\frac{ml^{2}}{\hbar}\tau\right)
,\ \label{1.4}%
\end{equation}
so that $\left\vert \Psi\left(  x,t\right)  \right\vert ^{2}dx=\left\vert
\psi\left(  q,\tau\right)  \right\vert ^{2}dq$.

In the new variables, equation (\ref{1.2}) takes the form%
\begin{equation}
\hat{S}\psi\left(  q,\tau\right)  =0,\text{ \ }\hat{S}=i\partial_{\tau}%
-\hat{H}, \label{1.5}%
\end{equation}
where the new Hamiltonian reads
\begin{equation}
\hat{H}=\frac{\hat{p}^{2}}{2}+\alpha\hat{q}^{2}+\beta\left(  \hat{q}\hat
{p}+\hat{p}\hat{q}\right)  +\varrho\hat{q}+\nu\hat{p}+\varepsilon\,.
\label{1.6}%
\end{equation}
Here $\alpha=\alpha\left(  \tau\right)  $, $\beta=\beta\left(  \tau\right)  $,
$\varrho=\varrho\left(  \tau\right)  $, $\nu=\nu\left(  \tau\right)  $ and
$\varepsilon=\varepsilon\left(  \tau\right)  ,$
\begin{align}
\alpha\left(  \tau\right)   &  =\frac{l^{4}}{2\hbar^{2}}\frac{r_{2}\left(
t\right)  }{r_{1}\left(  t\right)  },\ \beta\left(  \tau\right)  =\frac{l^{2}%
}{2\hbar}\frac{r_{3}\left(  t\right)  }{r_{1}\left(  t\right)  }%
,\ \ \varrho\left(  \tau\right)  =\frac{l^{3}}{2\hbar^{2}}\frac{r_{4}\left(
t\right)  }{r_{1}\left(  t\right)  },\nonumber\\
\nu\left(  \tau\right)   &  =\frac{l}{2\hbar}\frac{r_{5}\left(  t\right)
}{r_{1}\left(  t\right)  },\ \ \varepsilon(\tau)=\frac{l^{2}}{2\hbar^{2}}%
\frac{r_{6}\left(  t\right)  }{r_{1}\left(  t\right)  }\,, \label{2.5}%
\end{align}
are dimensionless real functions on $\tau$ if $t$ is expressed via $\tau$ by
the help of eqs. (\ref{1.3}). In what follows, we call $\hat{S}$ the equation operator.

\section{Constructing time-dependent generalized CS}

\subsection{Integrals of motion linear in canonical operators $\hat{q}$ and
$\hat{p}$\label{S3}}

First we construct an integral of motion $\hat{A}\left(  \tau\right)  $ linear
in $\hat{q}$ and $\hat{p}$. The general form of such an integral of motion
reads
\begin{equation}
\hat{A}\left(  \tau\right)  =f\left(  \tau\right)  \hat{q}+ig\left(
\tau\right)  \hat{p}+\varphi\left(  \tau\right)  , \label{3.1}%
\end{equation}
where $f\left(  \tau\right)  $, $g\left(  \tau\right)  $ and $\varphi\left(
\tau\right)  $ are some complex functions on $\tau$. The operator $\hat
{A}\left(  \tau\right)  $ is an integral of motion if it commutes with
equation operator (\ref{1.5}),
\begin{equation}
\left[  \hat{S},\hat{A}\left(  \tau\right)  \right]  =0. \label{3.2}%
\end{equation}
In the case if the Hamiltonian is self-adjoint, the adjoint operator $\hat
{A}^{\dag}\left(  \tau\right)  $ is also an integral of motion, i.e.,
\begin{equation}
\left[  \hat{S},\hat{A}^{\dag}\left(  \tau\right)  \right]  =0. \label{3.3}%
\end{equation}
The commutator $\left[  \hat{A}\left(  \tau\right)  ,\hat{A}^{\dag}\left(
\tau\right)  \right]  $ reads
\begin{equation}
\left[  \hat{A}\left(  \tau\right)  ,\hat{A}^{\dag}\left(  \tau\right)
\right]  =\delta=2\operatorname{Re}\left[  g^{\ast}\left(  \tau\right)
f\left(  \tau\right)  \right]  . \label{3.4}%
\end{equation}

Substituting representation (\ref{3.1}) into (\ref{3.2}), we obtain the
following equations for the functions $f\left(  \tau\right)  $, $g\left(
\tau\right)  ,$ and $\varphi\left(  \tau\right)  $:
\begin{align}
&  \dot{f}\left(  \tau\right)  +2\beta\left(  \tau\right)  f\left(
\tau\right)  -2i\alpha\left(  \tau\right)  g\left(  \tau\right)
=0,\nonumber\\
&  \dot{g}\left(  \tau\right)  -if\left(  \tau\right)  -2\beta\left(
\tau\right)  g\left(  \tau\right)  =0,\nonumber\\
&  \dot{\varphi}\left(  \tau\right)  +\nu\left(  \tau\right)  f\left(
\tau\right)  -i\varrho\left(  \tau\right)  g\left(  \tau\right)  =0.
\label{3.5}%
\end{align}
It is enough to find the functions $f\left(  \tau\right)  $ and $g\left(
\tau\right)  ,$ then the function $\varphi\left(  \tau\right)  $ can be found
by a simple integration. In addition, without loss of the generality we can
set $\varphi\left(  0\right)  =0.$

Equations (\ref{3.5}) imply that $\delta$ is a real integral of motion,
$\delta=\mathrm{const}.$ In what follows we suppose that $\delta=1,$ which
means%
\begin{equation}
\operatorname{Re}\left[  g^{\ast}\left(  \tau\right)  f\left(  \tau\right)
\right]  =\operatorname{Re}\left[  g^{\ast}\left(  0\right)  f\left(
0\right)  \right]  =1/2. \label{3.6}%
\end{equation}

Any nontrivial solution of two first equations (\ref{3.5}) consists of two
nonzero functions $f\left(  \tau\right)  $ and $g\left(  \tau\right)  .$ That
is why we can chose arbitrary integration constants in these equations as%
\begin{equation}
f\left(  0\right)  =c_{1}=\left\vert c_{1}\right\vert e^{i\mu_{1}%
},\ \ g\left(  0\right)  =c_{2}=\left\vert c_{2}\right\vert e^{i\mu_{2}%
},\ \ \left\vert c_{2}\right\vert \neq0,\ \ \left\vert c_{1}\right\vert \neq0.
\label{3.7}%
\end{equation}
In terms of the introduced constants, condition (\ref{3.6}) yields%
\begin{equation}
\left\vert c_{2}\right\vert \left\vert c_{1}\right\vert \cos\left(  \mu
_{1}-\mu_{2}\right)  =1/2. \label{3.8}%
\end{equation}

Under the choice $\delta=1,$ operators $\hat{A}\left(  \tau\right)  $ and
$\hat{A}^{\dag}\left(  \tau\right)  $ become annihilation and creation
operators,%
\begin{equation}
\left[  \hat{A}\left(  \tau\right)  ,\hat{A}^{\dag}\left(  \tau\right)
\right]  =1. \label{3.9}%
\end{equation}

It follows from eq. (\ref{3.1}) and (\ref{3.6}) that%
\begin{align}
&  \hat{q}=g^{\ast}\left(  \tau\right)  \left[  \hat{A}\left(  \tau\right)
-\varphi\left(  \tau\right)  \right]  +g\left(  \tau\right)  \left[  \hat
{A}^{\dag}\left(  \tau\right)  -\varphi^{\ast}\left(  \tau\right)  \right]
,\nonumber\\
&  i\hat{p}=f^{\ast}\left(  \tau\right)  \left[  \hat{A}\left(  \tau\right)
-\varphi\left(  \tau\right)  \right]  -f\left(  \tau\right)  \left[  \hat
{A}^{\dag}\left(  \tau\right)  -\varphi^{\ast}\left(  \tau\right)  \right]  .
\label{3.10}%
\end{align}

I. We note that the two first equations (\ref{3.5}) can be reduced to a one
second-order differential equation for the function $g\left(  \tau\right)  ,$
such an equation has the form of the oscillator equation with a time-dependent
frequency $\omega^{2}\left(  \tau\right)  $,
\begin{equation}
\ddot{g}\left(  \tau\right)  +\omega^{2}\left(  \tau\right)  g\left(
\tau\right)  =0,\text{ \ }\omega^{2}\left(  \tau\right)  =2\alpha-4\beta
^{2}-2\dot{\beta}. \label{3.12}%
\end{equation}
If we have an exact solution $g\left(  \tau\right)  $ for a given function
$\omega^{2}\left(  \tau\right)  ,$ then the function $f\left(  \tau\right)  $
can be found via the function $g\left(  \tau\right)  $ as
\begin{equation}
f\left(  \tau\right)  =2i\beta\left(  \tau\right)  g\left(  \tau\right)
-i\dot{g}\left(  \tau\right)  . \label{3.13}%
\end{equation}
One can chose the functions $\alpha\left(  \tau\right)  $ and $\beta\left(
\tau\right)  $ such that%
\begin{equation}
\text{\ }\omega^{2}\left(  \tau\right)  =2\alpha\left(  \tau\right)
-4\beta^{2}\left(  \tau\right)  -2\dot{\beta}\left(  \tau\right)  .
\label{3.14}%
\end{equation}
For example, if we chose
\begin{equation}
\alpha\left(  \tau\right)  =\frac{1}{2}\omega^{2}\left(  \tau\right)
,\ \beta=\varrho=\nu=\varepsilon=0, \label{3.13a}%
\end{equation}
then we are dealing with Hamiltonian of the form%
\begin{equation}
\hat{H}=\frac{\hat{p}^{2}}{2}+\frac{\omega^{2}\left(  \tau\right)  }{2}q^{2}.
\label{3.13b}%
\end{equation}

II. In addition, the one-dimensional Schrödinger equation%
\begin{equation}
-d_{q}^{2}\Psi\left(  q\right)  +V\left(  q\right)  \Psi\left(  q\right)
=E\Psi\left(  q\right)  , \label{u6}%
\end{equation}
can be identified with eq. (\ref{3.12}) if $q\rightarrow\tau,\ \Psi\left(
q\right)  \rightarrow g\left(  \tau\right)  ,\ V\left(  q\right)
-E\rightarrow\omega^{2}\left(  \tau\right)  .$

III. It should be also noted that two first equations (\ref{3.5}) can be
identified with a particular form of the so-called spin equation, see
\cite{BagBaGL},%
\begin{equation}
i\dot{V}=\left(  \boldsymbol{\sigma}\mathbf{F}\right)  V,\ \ V=\left(
\begin{array}
[c]{c}%
f\\
g
\end{array}
\right)  , \label{3.11}%
\end{equation}
with
\[
\mathbf{F}\left(  \tau\right)  =-\frac{1}{2}\left(  2\alpha+1,i\left(
2\alpha-1\right)  ,4i\beta\right)  .
\]

\subsection{Time-dependent generalized CS\label{S4}}

Let us consider eigenvectors $\left\vert z,\tau\right\rangle $ of the
annihilation operator $\hat{A}\left(  \tau\right)  $ corresponding to the
eigenvalue $z,$
\begin{equation}
\hat{A}\left(  \tau\right)  \left\vert z,\tau\right\rangle =z\left\vert
z,\tau\right\rangle . \label{4.1}%
\end{equation}
In the general case $z$ is a complex number.

It follows from eqs. (\ref{3.10}) and (\ref{4.1}) that%
\begin{align}
&  q\left(  \tau\right)  \equiv\left\langle z,\tau\left\vert \hat
{q}\right\vert z,\tau\right\rangle =g^{\ast}\left(  \tau\right)  \left[
z-\varphi\left(  \tau\right)  \right]  +g\left(  \tau\right)  \left[  z^{\ast
}-\varphi^{\ast}\left(  \tau\right)  \right]  ,\nonumber\\
&  ip\left(  \tau\right)  \equiv\left\langle z,\tau\left\vert \hat
{p}\right\vert z,\tau\right\rangle =f^{\ast}\left(  \tau\right)  \left[
z-\varphi\left(  \tau\right)  \right]  -f\left(  \tau\right)  \left[  z^{\ast
}-\varphi^{\ast}\left(  \tau\right)  \right]  ,\nonumber\\
&  z=f\left(  \tau\right)  q\left(  \tau\right)  +ig\left(  \tau\right)
p\left(  \tau\right)  +\varphi\left(  \tau\right)  . \label{4.2}%
\end{align}

Using (\ref{3.5}), one can easily verify that the functions $q\left(
\tau\right)  $ and $p\left(  \tau\right)  $ satisfy the Hamilton equations%
\[
\dot{q}\left(  \tau\right)  =\frac{\partial H}{\partial p},\quad\dot{p}\left(
\tau\right)  =-\frac{\partial H}{\partial q}\,,
\]
where $H=H\left(  q,p\right)  $ is the classical Hamiltonian that corresponds
to the quantum Hamiltonian (\ref{1.6}). Thus, the pair $q\left(  \tau\right)
$ and $p\left(  \tau\right)  $ represents a classical trajectory in the phase
space of the system under consideration. All such trajectories can be
parametrized by the initial data, $q_{0}=q\left(  0\right)  $ and\ $p_{0}%
=p\left(  0\right)  .$

Being written in the $q$-representation, eq. (\ref{4.1}) reads%
\begin{equation}
\left[  f\left(  \tau\right)  q+g\left(  \tau\right)  \partial_{q}%
+\varphi\left(  \tau\right)  \right]  \langle q\left\vert z,\tau\right\rangle
=z\langle q\left\vert z,\tau\right\rangle . \label{4.4}%
\end{equation}
General solution of this equation has the form%
\begin{equation}
\langle q\left\vert z,\tau\right\rangle =\Phi_{z}^{c_{1}c_{2}}\left(
q,\tau\right)  =\exp\left[  -\frac{f\left(  \tau\right)  }{g\left(
\tau\right)  }\frac{q^{2}}{2}+\frac{z-\varphi\left(  \tau\right)  }{g\left(
\tau\right)  }q+\chi\left(  \tau\right)  \right]  , \label{4.5}%
\end{equation}
where $\chi\left(  \tau\right)  $ is an arbitrary function on $\tau$.

One can see that the functions $\Phi_{z}^{c_{1}c_{2}}\left(  q,\tau\right)  $
can be written in terms of the mean values $q\left(  \tau\right)  $ and
$p\left(  \tau\right)  $ given by eqs. (\ref{4.2}),%
\begin{equation}
\Phi_{z}^{c_{1}c_{2}}\left(  q,\tau\right)  =\exp\left\{  ip\left(
\tau\right)  q-\frac{f\left(  \tau\right)  }{2g\left(  \tau\right)  }\left[
q-q\left(  \tau\right)  \right]  ^{2}+\tilde{\chi}\left(  \tau\right)
\right\}  . \label{4.6}%
\end{equation}
where $\tilde{\chi}\left(  \tau\right)  $ is again an arbitrary function on
$\tau$.

The functions $\Phi_{z}^{c_{1}c_{2}}$ satisfy the following equation%
\begin{equation}
\hat{S}\Phi_{z}^{c_{1}c_{2}}\left(  q,\tau\right)  =\lambda\left(
\tau\right)  \Phi_{z}^{c_{1}c_{2}}\left(  q,\tau\right)  , \label{4.7}%
\end{equation}
where%
\begin{equation}
\lambda\left(  \tau\right)  =i\partial_{\tau}\tilde{\chi}\left(  \tau\right)
+\alpha q^{2}\left(  \tau\right)  -\frac{1}{2}\left[  p^{2}\left(
\tau\right)  +\frac{f}{g}\right]  -i\nu p\left(  \tau\right)  -\beta
-i\varepsilon. \label{4.8}%
\end{equation}

If we wish the functions (\ref{4.6}) satisfies the Schrödinger equation
(\ref{1.5}), we have to fix $\tilde{\chi}\left(  \tau\right)  $ from the
condition $\lambda\left(  \tau\right)  =0$. Thus, we obtain for the function
$\tilde{\chi}\left(  \tau\right)  $ the following result:%
\begin{align}
\tilde{\chi}\left(  \tau\right)   &  =\phi\left(  \tau\right)  +\ln
N,\nonumber\\
\phi\left(  \tau\right)   &  =\int_{0}^{\tau}\left\{  i\alpha q^{2}\left(
\tau\right)  -\frac{i}{2}\left[  p^{2}\left(  \tau\right)  +\frac{f}%
{g}\right]  -i\nu p\left(  \tau\right)  -\beta-i\varepsilon\right\}  d\tau,
\label{4.9}%
\end{align}
were $N$ is a normalization constant, which we suppose to be real.

The probability densities generated by the wave functions (\ref{4.6}) have the
form%
\begin{equation}
\rho_{z}^{c_{1}c_{2}}\left(  q,\tau\right)  =\left\vert \Phi_{z}^{c_{1}c_{2}%
}\left(  q,\tau\right)  \right\vert ^{2}=N^{2}\exp\left\{  -\frac{\left[
q-q\left(  \tau\right)  \right]  ^{2}}{2\left\vert g\left(  \tau\right)
\right\vert ^{2}}+2\operatorname{Re}\phi\left(  \tau\right)  \right\}  .
\label{4.10}%
\end{equation}
Considering the normalization integral, we find the constant $N$,%
\begin{equation}
\int_{-\infty}^{\infty}\rho_{z}^{c_{1}c_{2}}\left(  q,\tau\right)
dq=1\Rightarrow N=\frac{\exp\left(  -\operatorname{Re}\phi\left(  \tau\right)
\right)  }{\sqrt{\sqrt{2\pi}\left\vert g\left(  \tau\right)  \right\vert }}.
\label{4.11}%
\end{equation}

Thus, normalized solutions of the Schrödinger equation that at the same time
are eigenfunctions of the annihilation operator $\hat{A}\left(  \tau\right)  $
have the form%
\begin{equation}
\Phi_{z}^{c_{1}c_{2}}\left(  q,\tau\right)  =\frac{1}{\sqrt{\sqrt{2\pi
}\left\vert g\left(  \tau\right)  \right\vert }}\exp\left\{  ip\left(
\tau\right)  q-\frac{f\left(  \tau\right)  }{g\left(  \tau\right)  }%
\frac{\left[  q-q\left(  \tau\right)  \right]  ^{2}}{2}+i\operatorname{Im}%
\phi\left(  \tau\right)  \right\}  \label{4.12}%
\end{equation}
and the corresponding probability densities read%
\begin{equation}
\rho_{z}^{c_{1}c_{2}}\left(  q,\tau\right)  =\frac{1}{\sqrt{2\pi}\left\vert
g\left(  \tau\right)  \right\vert }\exp\left\{  -\frac{\left[  q-q\left(
\tau\right)  \right]  ^{2}}{2\left\vert g\left(  \tau\right)  \right\vert
^{2}}\right\}  . \label{4.13}%
\end{equation}

In what follows we call the solutions (\ref{4.12}) the time-dependent
generalized CS.

\section{Time-dependent CS of quadratic systems}

Using eqs. (\ref{3.10}) and (\ref{4.1}) we can calculate standard deviations
$\sigma_{q}\left(  \tau\right)  $, $\sigma_{p}\left(  \tau\right)  $, and the
quantity $\sigma_{qp}\left(  \tau\right)  $, in the generalized CS,%
\begin{align}
&  \sigma_{q}\left(  \tau\right)  =\sqrt{\langle\left(  \hat{q}-\left\langle
q\right\rangle \right)  ^{2}\rangle}=\sqrt{\left\langle q^{2}\right\rangle
-\left\langle q\right\rangle ^{2}}=\left\vert g\left(  \tau\right)
\right\vert ,\nonumber\\
&  \sigma_{p}\left(  \tau\right)  =\sqrt{\langle\left(  \hat{p}-\left\langle
p\right\rangle \right)  ^{2}\rangle}=\sqrt{\left\langle p^{2}\right\rangle
-\left\langle p\right\rangle ^{2}}=\left\vert f\left(  \tau\right)
\right\vert ,\nonumber\\
&  \sigma_{qp}\left(  \tau\right)  =\frac{1}{2}\left\langle \left(  \hat
{q}-\left\langle q\right\rangle \right)  \left(  \hat{p}-\left\langle
p\right\rangle \right)  +\left(  \hat{p}-\left\langle p\right\rangle \right)
\left(  \hat{q}-\left\langle q\right\rangle \right)  \right\rangle \nonumber\\
&  =i\left[  1/2-g\left(  \tau\right)  f^{\ast}\left(  \tau\right)  \right]  .
\label{5.1}%
\end{align}

One can easily see that the generalized CS (\ref{4.12}) minimize the
Robertson-Schrödinger uncertainty relation \cite{SchroRo},%
\begin{equation}
\sigma_{q}^{2}\left(  \tau\right)  \sigma_{p}^{2}-\sigma_{qp}^{2}\left(
\tau\right)  =1/4. \label{5.2}%
\end{equation}
This means that the generalized CS are squeezed states \cite{DodMa87}.

Let us analyze the Heisenberg uncertainty relation in the generalized CS
taking into account restriction (\ref{3.6}),%
\begin{equation}
\left.  \sigma_{q}\left(  \tau\right)  \sigma_{p}\left(  \tau\right)
\right\vert _{2\operatorname{Re}\left(  c_{1}^{\ast}c_{2}\right)  }=\frac
{1}{2}\sqrt{1+4\left(  \operatorname{Im}\left(  gf^{\ast}\right)  \right)
^{2}}\geq\frac{1}{2}. \label{5.3}%
\end{equation}

Then using (\ref{5.1}), we find $\sigma_{q}\left(  0\right)  =\sigma
_{q}=\left\vert c_{2}\right\vert $ and $\sigma_{p}\left(  0\right)
=\sigma_{p}=\left\vert c_{1}\right\vert ,$ such that at $\tau=0$ this relation
reads
\begin{equation}
\left.  \sigma_{q}\sigma_{p}\right\vert _{2\operatorname{Re}\left(
c_{1}^{\ast}c_{2}\right)  }=\sqrt{\frac{1}{4}+\left[  \left\vert
c_{2}\right\vert \left\vert c_{1}\right\vert \sin\left(  \mu_{2}-\mu
_{1}\right)  \right]  ^{2}}. \label{5.4}%
\end{equation}
Taking into account eqs. (\ref{3.7}), we see that if $\mu_{1}=\mu_{2}=\mu$ the
left hand side of (\ref{5.3}) is minimal, such that%
\begin{equation}
\sigma_{q}\sigma_{p}=1/2,\ \ \sigma_{qp}=0. \label{5.5}%
\end{equation}

One can see that the constant $\mu$ does not enter CS (\ref{4.12}). Then, in
what follows we consider generalized CS with the restriction $\mu_{1}=\mu
_{2}=\mu=0$. Namely, such states we call simply CS.

Now restriction (\ref{3.6}) takes the form $c_{1}=\left\vert c_{1}\right\vert
,\ c_{2}=\left\vert c_{2}\right\vert ,\ 2c_{1}=c_{2}^{-1},$ such that%
\begin{equation}
g\left(  0\right)  =\left\vert c_{2}\right\vert =\sigma_{q},\ \ f\left(
0\right)  =\left\vert c_{1}\right\vert =\sigma_{p}=\frac{1}{2\sigma_{q}}.
\label{5.8}%
\end{equation}
Thus, $\sigma_{qp}=\sigma_{qp}\left(  0\right)  =i\left[  1/2-g\left(
0\right)  f\left(  0\right)  \right]  =0,$ which is consistent with eqs.
(\ref{5.5}).

With account taken of eqs. (\ref{4.12}), (\ref{5.1}) and (\ref{5.8}), we
obtain the following expression for the CS:%
\begin{align}
&  \Phi_{z}^{\sigma_{q}}\left(  q,\tau\right)  =\frac{1}{\sqrt{\sqrt{2\pi
}\sigma_{q}\left(  \tau\right)  }}\exp\left\{  ip\left(  \tau\right)
q-\frac{f\left(  \tau\right)  }{g\left(  \tau\right)  }\frac{\left[
q-q\left(  \tau\right)  \right]  ^{2}}{2}+i\operatorname{Im}\phi\left(
\tau\right)  \right\}  ,\nonumber\\
&  \phi\left(  \tau\right)  =\int_{0}^{\tau}\left\{  i\alpha q^{2}\left(
\tau\right)  -\frac{i}{2}\left[  p^{2}\left(  \tau\right)  +\frac{f\left(
\tau\right)  }{g\left(  \tau\right)  }\right]  -i\nu p\left(  \tau\right)
-\beta-i\varepsilon\right\}  d\tau. \label{5.9}%
\end{align}

In fact, we have a family of CS parametrized by one real parameter-the initial
standard deviation $\sigma_{q}>0$. Each set of CS in the family has its
specific initial standard deviations $\sigma_{q}$. Different CS from a family
with a given $\sigma_{q}$ have different quantum numbers $z,$ which are in one
to one correspondence with trajectory initial data $q_{0}$ and $p_{0}.$ It
follows from eq. (\ref{4.2}) that%
\begin{equation}
z=\frac{q_{0}}{2\sigma_{q}}+i\sigma_{q}p_{0},\ \ q_{0}=2\sigma_{q}%
\operatorname{Re}z\ ,\ \ p_{0}=\frac{\operatorname{Im}z}{\sigma_{q}}\ .
\label{5.10a}%
\end{equation}
The probability density that corresponds to the CS (\ref{5.9}) reads%
\begin{equation}
\rho_{z}^{\sigma_{q}}\left(  q,\tau\right)  =\frac{1}{\sqrt{2\pi}\sigma
_{q}\left(  \tau\right)  }\exp\left\{  -\frac{\left[  q-q\left(  \tau\right)
\right]  ^{2}}{2\sigma_{q}^{2}\left(  \tau\right)  }\right\}  . \label{5.11}%
\end{equation}

One can prove that for any fixed $\sigma_{q}$ states (\ref{5.9}) form an over
complete set of functions with the following orthogonality and completeness
relations%
\begin{align}
&  \int\overline{\Phi_{z^{\prime}}^{\sigma_{q}}\left(  q,\tau\right)  }%
\Phi_{z}^{\sigma_{q}}\left(  q,\tau\right)  dq=\exp\left(  z^{\prime\ast
}z-\frac{\left\vert z^{\prime}\right\vert ^{2}+\left\vert z\right\vert ^{2}%
}{2}\right)  ,\ \ \forall\tau,\nonumber\\
&  \int\int\Phi_{z}^{\sigma_{q}}\left(  q,\tau\right)  \overline{\Phi
_{z}^{\sigma_{q}}\left(  q^{\prime},\tau\right)  }d^{2}z=\pi\delta\left(
q-q^{\prime}\right)  ,\text{ \ }d^{2}z=d\operatorname{Re}z\ d\operatorname{Im}%
z,\text{ \ }\forall\tau. \label{5.12}%
\end{align}

\section{An exact solution of oscillator equation with time-dependent
frequency and related CS}

Let us consider the following function $\omega^{2}\left(  \tau\right)  ,$%
\begin{equation}
\omega^{2}\left(  \tau\right)  =\omega^{2}+\frac{2\omega_{0}^{2}}{\cosh
^{2}\omega_{0}\tau},\text{ \ }\omega^{2}\leq\omega^{2}\left(  \tau\right)
\leq\omega_{\max}^{2},\text{ \ }\omega\left(  \pm\infty\right)  =\omega
^{2},\label{6.1}%
\end{equation}
where $\omega$ and $\omega_{0}$ are some positive constants,\ $\omega_{\max
}\geq$ $\omega$. The function $\omega^{2}\left(  \tau\right)  $ is an even
function, which decreases monotonically as $|\tau|$ changes from $0$ to
$\infty,$%
\begin{equation}
\omega^{2}\left(  \pm\infty\right)  =\omega^{2}\ ,\ \ \omega^{2}\left(
0\right)  =\omega_{\max}^{2}=\omega^{2}+2\omega_{0}^{2}\ .\label{6.2}%
\end{equation}
For $\omega=1$ and $\omega_{0}=2^{-1/2}$, the plot of the function
$\omega\left(  \tau\right)  $ has the form%

\begin{figure}[th]
\begin{center}
\includegraphics[scale=1]{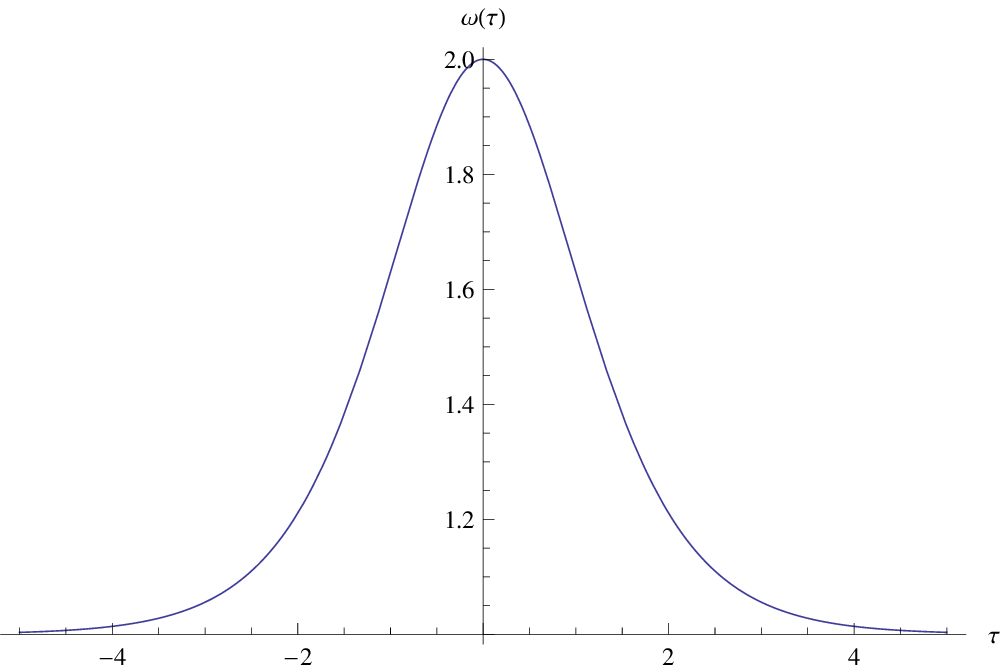}
\end{center}
\label{fig1}
\end{figure}

The general solution of equation (\ref{3.12}) can be written as
\begin{equation}
g(\tau)=\left[  \frac{iA\,\omega_{0}\tanh(\omega_{0}\tau)}{\omega^{2}%
+\omega_{0}^{2}}+B\right]  \cos(\omega\tau)+\left[  \frac{iA\,\omega^{2}%
}{\omega^{2}+\omega_{0}^{2}}-B\,\omega_{0}\tanh(\omega_{0}\tau)\right]
\frac{\sin(\omega\tau)}{\omega}. \label{6.4}%
\end{equation}

The restriction (\ref{3.6}), (\ref{3.7}), and (\ref{5.8}) that set the CS from
the entire set of generalized CS lead to the following relations for the
constants $A$ and $B$:%
\begin{equation}
B=g\left(  0\right)  =\left\vert g\left(  0\right)  \right\vert =\sigma
_{q},\ \ A=f(0)=\left\vert f\left(  0\right)  \right\vert =1/2\sigma_{q}.
\label{6.5}%
\end{equation}

Using eqs. (\ref{6.4}) and (\ref{6.5}) we calculate the mean trajectories
$q(\tau)$ and $p\left(  \tau\right)  $ according eqs. (\ref{4.2}),%
\begin{align}
&  q(\tau)=2\operatorname{Re}\left[  g\left(  \tau\right)  z^{\ast}\right]
=\left.  g\left(  \tau\right)  \right\vert _{A=\bar{A},\ B=\bar{B}%
}\ ,\nonumber\\
&  p(\tau)=\left.  \dot{g}\left(  \tau\right)  \right\vert _{A=\bar
{A},\ B=\bar{B}}=\dot{q}(\tau)\,,\ \overline{A}=-\sigma_{q}^{-1}%
\operatorname{Im}z\,,\ \ \overline{B}=2\sigma_{q}\operatorname{Re}z\,.
\label{6.6}%
\end{align}

For $\omega_{\min}>0$ the mean trajectory $q(\tau)$ can be presented as
\begin{equation}
q(\tau)=R_{0}R(\tau)\sin[\omega\tau+\Theta(\tau)+\Theta_{0}]\,, \label{6.11}%
\end{equation}
where functions $R(\tau)$ and $\Theta(\tau)$ and constants $R_{0}$ and
$\Theta_{0}$ are%
\begin{align}
&  R(\tau)=\sqrt{1+\frac{\omega_{0}^{2}}{\omega^{2}}\tanh^{2}\omega_{0}\tau
}\,,\ \ 1\leqslant R(\tau)<\omega^{-1}\sqrt{\omega^{2}+\omega_{0}^{2}%
}\,;\nonumber\\
&  \Theta(\tau)=\arctan\left[  \frac{\omega_{0}}{\omega}\tanh\omega_{0}%
\tau\right]  ,\ \ -\Delta<\Theta(\tau)<\Delta\,,\ \ \Delta=\arctan\left(
\frac{\omega_{0}}{\omega}\right)  \,;\nonumber\\
&  R_{0}=\frac{\sqrt{p_{0}^{2}\omega^{2}+q_{0}^{2}(\omega^{2}+\omega_{0}%
^{2})^{4}}}{(\omega^{2}+\omega_{0}^{2})^{2}},\ \ \sin\Theta_{0}=\frac{q_{0}%
}{R_{0}},\ \ \cos\Theta_{0}=\frac{p_{0}\,\omega}{R_{0}(\omega^{2}+\omega
_{0}^{2})^{2}}\,. \label{6.13}%
\end{align}
Thus, we deal with a quasiharmonic motion with the frequency $\omega$\ and an
amplitude that is changing in time in finite limits and with a time-dependent
phase that is slowly changing in also finite limits.

Let us derive the case of a harmonic oscillator with a fixed frequency
$\omega>0$ from the above formulas. To this end we have to set $\omega_{0}=0$
and $\alpha=\omega^{2}/2,$ $\nu=\beta=\varepsilon=0$ such that $\omega
^{2}\left(  \tau\right)  =\omega^{2}$. Then%
\begin{align}
g\left(  \tau\right)   &  =\sigma_{q}\cos\omega\tau+\frac{i\sin\omega\tau
}{2\sigma_{q}\omega},\ \ f\left(  \tau\right)  =\frac{\cos\omega\tau}%
{2\sigma_{q}}+i\sigma_{q}\omega\sin\omega\tau,\nonumber\\
q(\tau)  &  =q_{0}\cos\omega\tau+\frac{p_{0}}{\omega}\sin\omega\tau
,\ p(\tau)=p_{0}\cos\omega\tau-\omega q_{0}\sin\omega\tau, \label{6.14}%
\end{align}
and $z=\sigma_{p}q_{0}+i\sigma_{q}\,p_{0}\,.$ Taking all that into account in
Eq. (\ref{5.9}), we obtain the following representation for CS (in the above
given definition)\ of the harmonic oscillator:%
\begin{equation}
\Phi_{z}^{\sigma_{q}}\left(  q,\tau\right)  =\frac{1}{\sqrt{\sqrt{2\pi
}g\left(  \tau\right)  }}\exp\left\{  -\frac{1}{2}\frac{f\left(  \tau\right)
}{g\left(  \tau\right)  }\left[  q-\frac{z}{f\left(  \tau\right)  }\right]
^{2}+\frac{f^{\ast}\left(  \tau\right)  }{f\left(  \tau\right)  }\frac{z^{2}%
}{2}-\frac{\left\vert z\right\vert ^{2}}{2}\right\}  . \label{6.17}%
\end{equation}
For these CS
\begin{equation}
\sigma_{q}\left(  \tau\right)  =\sigma_{q}\sqrt{1+\frac{\left(  1-4\sigma
_{q}^{4}\omega^{2}\right)  }{4\sigma_{q}^{4}\omega^{2}}\sin^{2}\omega\tau
},\ \ \sigma_{p}\left(  \tau\right)  =\sigma_{p}\sqrt{1-\left(  1-4\sigma
_{q}^{4}\omega^{2}\right)  \sin^{2}\omega\tau}, \label{6.15}%
\end{equation}
and the corresponding probability density reads%
\begin{equation}
\rho_{z}^{\sigma_{q}}\left(  q,\tau\right)  =\frac{1}{\sqrt{2\pi}\sigma
_{q}\left(  \tau\right)  }\exp\left\{  -\frac{\left[  q-q\left(  \tau\right)
\right]  ^{2}}{2\sigma_{q}^{2}\left(  \tau\right)  }\right\}  . \label{6.16}%
\end{equation}

One has to consider the following three cases:

a) $\sigma_{q}\sqrt{2\omega}=1,$ then%
\[
\sigma_{q}\left(  \tau\right)  =\sigma_{q},\ \ \sigma_{p}\left(  \tau\right)
=\sigma_{p},\ \ \sigma_{q}\left(  \tau\right)  \sigma_{p}\left(  \tau\right)
=\sigma_{q}\sigma_{p}=1/2,\ \forall\tau.
\]

b) $\sigma_{q}\sqrt{2\omega}<1,$ then%
\begin{align*}
&  \left.  \sigma_{q}\left(  \tau\right)  \right\vert _{\min}=\left.
\sigma_{q}\left(  \tau\right)  \right\vert _{\tau=\frac{\pi n}{\omega}}%
=\sigma_{q},\ \ \left.  \sigma_{q}\left(  \tau\right)  \right\vert _{\max
}=\left.  \sigma_{q}\left(  \tau\right)  \right\vert _{\tau=\frac{2n+1}%
{2}\frac{\pi}{\omega}}=\frac{1}{2\sigma_{q}\omega},\\
&  \left.  \sigma_{p}\left(  \tau\right)  \right\vert _{\min}=\left.
\sigma_{p}\left(  \tau\right)  \right\vert _{\tau=\frac{2n+1}{2}\frac{\pi
}{\omega}}=\sigma_{q}\omega,\ \ \left.  \sigma_{p}\left(  \tau\right)
\right\vert _{\max}=\left.  \sigma_{p}\left(  \tau\right)  \right\vert
_{\tau=\frac{\pi n}{\omega}}=\sigma_{p},\\
&  \left.  \sigma_{q}\left(  \tau\right)  \sigma_{p}\left(  \tau\right)
\right\vert _{\min}=\left.  \sigma_{q}\left(  \tau\right)  \sigma_{p}\left(
\tau\right)  \right\vert _{\tau=\frac{\pi n}{\omega}}=1/2,\ \\
&  \left.  \sigma_{q}\left(  \tau\right)  \sigma_{q}\left(  \tau\right)
\right\vert _{\max}=\left.  \sigma_{q}\left(  \tau\right)  \sigma_{q}\left(
\tau\right)  \right\vert _{\tau=\frac{2n+1}{4}\frac{\pi}{\omega}}%
=\frac{1+4\sigma_{q}^{4}\omega^{2}}{8\sigma_{q}^{2}\omega},\ \ n\in
\mathbb{N}=0,1,2,...\ .
\end{align*}

c) $\sigma_{q}\sqrt{2\omega}>1,$ then
\begin{align*}
&  \left.  \sigma_{q}\left(  \tau\right)  \right\vert _{\min}=\left.
\sigma_{q}\left(  \tau\right)  \right\vert _{\tau=\frac{2n+1}{2}\frac{\pi
}{\omega}}=\frac{1}{2\sigma_{q}\omega},\ \ \left.  \sigma_{q}\left(
\tau\right)  \right\vert _{\max}=\left.  \sigma_{q}\left(  \tau\right)
\right\vert _{\tau=\frac{n\pi}{\omega}}=\sigma_{q},\\
&  \left.  \sigma_{p}\left(  \tau\right)  \right\vert _{\min}=\left.
\sigma_{p}\left(  \tau\right)  \right\vert _{\tau=\frac{n\pi}{\omega}}%
=\sigma_{p},\ \left.  \sigma_{p}\left(  \tau\right)  \right\vert _{\max
}=\left.  \sigma_{p}\left(  \tau\right)  \right\vert _{\tau=\frac{2n+1}%
{2}\frac{\pi}{\omega}}=\sigma_{q}\omega,\\
&  \left.  \sigma_{q}\left(  \tau\right)  \sigma_{p}\left(  \tau\right)
\right\vert _{\min}=\left.  \sigma_{q}\left(  \tau\right)  \sigma_{p}\left(
\tau\right)  \right\vert _{\tau=\frac{n}{2}\frac{\pi}{\omega}}=1/2,\\
&  \left.  \sigma_{q}\left(  \tau\right)  \sigma_{q}\left(  \tau\right)
\right\vert _{\max}=\left.  \sigma_{q}\left(  \tau\right)  \sigma_{q}\left(
\tau\right)  \right\vert _{\tau=\frac{2n+1}{4}\frac{\pi}{\omega}}%
=\frac{1+4\sigma_{q}^{4}\omega^{2}}{8\sigma_{q}^{2}\omega},\ \ n\in
\mathbb{N}\ .
\end{align*}

We can see that in the case a) the Heisenberg uncertainty relation is
minimized in the CS (\ref{6.17}). In the same case these CS coincide (up to a
phase factor) with the well-known Schrödinger-Glauber CS \cite{Glauber}. CS
with $\sigma_{q}$ obeying b) and c) minimize the Heisenberg uncertainty
relation periodically, but the product $\sigma_{q}\left(  \tau\right)
\sigma_{q}\left(  \tau\right)  $ is always restricted by the limits $1/2$ and
$\frac{1+4\sigma_{q}^{4}\omega^{2}}{8\sigma_{q}^{2}\omega}.$

Setting $\omega_{0}=\omega=\alpha=\nu=\beta=\varepsilon=0,$ and taking into
account the limits%
\begin{align*}
\lim_{\omega_{0}\rightarrow0}g\left(  \tau\right)   &  =B\cos\left(
\omega\tau\right)  +\frac{iA}{\omega}\sin\left(  \omega\tau\right)
,\ \lim_{\omega_{0},\omega\rightarrow0}g\left(  \tau\right)  =B+iA\tau,\\
\lim_{\omega\rightarrow0}g\left(  \tau\right)   &  =(iA-B\omega_{0}^{2}%
\tau)\frac{\tanh(\omega_{0}\tau)}{\omega_{0}}+B,
\end{align*}
we obtain from (\ref{5.9}) CS of a free particle studied by us in the Ref.
\cite{BagGiP14}.

\section*{Acknowledgements}

Bagrov thanks FAPESP for support and IF USP for hospitality;

Gitman thanks CNPq and FAPESP for permanent support;

The work of Bagrov and Gitman is also partially supported by Tomsk State
University A. S. Competitiveness Improvement Program;

Pereira thanks FAPESP for support.


\begin{thebibliography}{99}                                                                                               %


\bibitem {CSQT72}Klauder J R, Sudarshan E C, \emph{Fundamentals of Quantum
Optics}, (Benjamin, 1968); Klauder I R, Skagerstam B S, \emph{Coherent States,
Applications in Physics and Mathematical Physics}, (World Scientific,
Singapore, 1985); Perelomov A M, \emph{Generalized Coherent States and Their
Applications}, (Springer-Verlag, 1986); Gazeau J P, \emph{Coherent States in
Quantum Physics}, (Wiley-VCH, Berlin, 2009); Nielsen M, Chuang I,
\emph{Quantum Computation and Quantum Information} (Cambridge University
Press, Cambridge, England, 2000)

\bibitem {MalMa68}Malkin I A and Man'ko V I 1968 Zh. Eksp. Teor. Fiz.
\textbf{55} 1014

\bibitem {MalMa}Malkin I A, Man'ko V I, \emph{Dynamical Symmetries and
Coherent States of Quantum Systems}, (Nauka, Moscow, 1979)

\bibitem {DodMa87}V.V. Dodonov, V.I. Man'ko,{\small \ }\emph{Invariants and
correlated states of nonstationary quantum systems}. In: \textbf{Invariants
and the Evolution of Nonstationary Quantum Systems}. Proceedings of Lebedev
Physics Institute, \textbf{183,}\/M.A. Markov, ed. (Nauka, Moscow 1987)
71-181, [translated by Nova Science, Commack, New York, 1989, pp. 103-261

\bibitem {gilm72}R. Gilmore, \emph{Geometry of symmetrized states},
\textit{Ann. Phys. (NY)} \textbf{74} 391-463 (1972).

\bibitem {perel72}A. M. Perelomov, \emph{Coherent states for arbitrary Lie
groups}, \textit{Commun. Math. Phys.} \textbf{26} 222-236 (1972).

\bibitem {perel86}A.M.Perelomov, \emph{Generalized Coherent States and Their
Applications}, Springer-Verlag, 1986.

\bibitem {AAG00}S.T. Ali, J-P. Antoine, and J-P. Gazeau, \emph{Coherent
States, Wavelets and Their Generalizations}, Springer-Verlag, New York,
Berlin, Heidelberg, 2000.

\bibitem {DodMa03}\emph{Theory of Nonclassical States of Light, }Edited by V.
V. Dodonov and V. I. Man'ko (Taylor \& Francis Group, London, NY\ 2003)

\bibitem {16}V.G. Bagrov, I.L. Buchbinder and D.M. Gitman, \emph{Coherent
states of a relativistic particle in an external electromagnetic field},
Journ. Phys. A \textbf{9} (1976) 1955-1965

\bibitem {4}J.P. Gazeau, M.C. Baldiotti, and D.M. Gitman, \emph{Semiclassical
and quantum motion on non-commutative plane,} Physics Letters A, \textbf{373},
(43) (2009) 3937-3943

\bibitem {265}V.G. Bagrov, S.P. Gavrilov, D.M. Gitman, and D. P. Meira Filho,
\emph{Coherent states of non-relativistic electron in magnetic--solenoid
field, }Journ. Physics. A\textbf{43} (2010) 3540169 (10 pages); \emph{Coherent
and semiclassical states in magnetic field in the presence of the
Aharonov-Bohm solenoid}, J. Phys. A: Math. Theor. \textbf{44} (2011) 055301

\bibitem {287}V.G. Bagrov, S.P. Gavrilov, D.M. Gitman, and K. Gorska,
\emph{Completeness for coherent states in magnetic-solenoid field}, Journ.
Phys. A \textbf{45} (2012) 244008 (11pp)

\bibitem {290}V.G. Bagrov, J.-P. Gazeau, D. Gitman, and A. Levine,
\emph{Coherent states and related quantizations for unbounded motions, }Journ.
Phys. A \textbf{45} (2012) 125306

\bibitem {295}V.G. Bagrov, D. M. Gitman, E. S. Macedo, and A. S. Pereira,
\emph{Coherent states of inverse oscillator and related problems}, Journ.
Phys. A: Math. Theor. \textbf{46} (2013) 325305 (13pp);

\bibitem {303}V.G. Bagrov, D.M. Gitman, A. Pereira, \emph{Coherent and
semiclassical states of a free particle,} Uspekhi Fizicheskikh Nauk
\textbf{184} (9)\emph{\ }(2014) 961-966, Physics-Uspekhi 57 (9) (2014) 891-896

\bibitem {Glauber}E. Schrödinger, \emph{Der stetige Übergang von der Mikro-
zur Makro-mechanik}, Naturwissenschaften, Bd. \textbf{14}, H. 28 (1926)
664-666;\ R.J. Glauber, \emph{The quantum theory of optical coherence},
\emph{Phys. Rev.} \textbf{130} No. 6 (1963) 2529--2539; \emph{Coherent and
incoherent states of the radiation field}, Phys. Rev. \textbf{131 }No. 6
(1963) 2766-2788

\bibitem {BagGiP14}V.G. Bagrov, D.M. Gitman, A. Pereira, \emph{Coherent and
semiclassical states of a free particle,} Physics-Uspekhi 57 (9) (2014) 891-896

\bibitem {book2}Bagrov V G, Gitman D M, \emph{Exact Solutions of Relativistic
Wave Equations} (Kluwer Acad. Publisher, Boston 1990); \emph{Dirac Equation
and its Solutions} (de Gruyter, Boston, 2014) pp. 444

\bibitem {BagBaGL}V.G. Bagrov, M.C. Baldiotti, D.M. Gitman, A.D. Levin,
\emph{Spin equation and its solutions, }Ann. der Physik \textbf{14} [11-12]
(2005) 764-789

\bibitem {SchroRo}Schrödinger E,\textit{ Sitzungsberichte Preus. Acad. Wiss.,
Phys.-Math. Klasse,} \textbf{19} 296 (Berlin 1930) ; Robertson H P,
\textit{Phys. Rev.} \textbf{35} 667 (1930)
\end{thebibliography}
\end{document}